\documentclass[12pt,preprint]{aastex}

\usepackage{amssymb,amsmath} 

\newcommand\beq{\begin{equation}}
\newcommand\eeq{\end{equation}}
\newcommand\beqar{\begin{eqnarray}}
\newcommand\eeqar{\end{eqnarray}}

\newcommand{\cref}{C_{\rm ref}}

\begin{document}

\title{Non-equilibrium Chemistry of Dynamically Evolving Prestellar Cores:\\ II.  
Ionization and Magnetic Field}

\author{Konstantinos Tassis\altaffilmark{1,2}, 
Karen Willacy\altaffilmark{1},
Harold W. Yorke\altaffilmark{1},
\& Neal J. Turner\altaffilmark{1}}

\altaffiltext{1}{Jet Propulsion Laboratory, California Institute of Technology, Pasadena, CA 91109, USA}
\altaffiltext{2}{Current address: Max-Planck Institut f\"ur Radioastronomie, 53121 Bonn, Germany}

\begin{abstract}

We study the effect that non-equilibrium chemistry in dynamical models of collapsing molecular cloud cores has on measurements of the magnetic field in these cores, the degree of ionization, and the mean molecular weight of ions. 
We find that OH and CN, usually used in Zeeman observations of the line-of-sight magnetic field, have an abundance that decreases toward the center of the core much faster than the density increases. As a result, Zeeman observations tend to sample the outer layers of the core and consistently underestimate the core magnetic field.
The degree of ionization follows a complicated dependence on the number density at central densities up to $10^5 \, {\rm cm^{-3}}$ for magnetic models and $10^6 \, {\rm cm^{-3}}$ in non-magnetic models. At higher central densities the scaling approaches a power-law with a slope of -0.6 and a normalization which depends on the cosmic-ray ionization rate $\zeta$ and the temperature $T$ as $(\zeta T)^{1/2}$. The mean molecular weight of ions is systematically lower than the usually assumed value of $20-30$, and, at high densities, approaches a value of 3 due to the asymptotic dominance of the H$_3^+$ ion. This significantly lower value implies that ambipolar diffusion operates faster. 

\end{abstract}


\section{Introduction}

Prestellar cores are the earliest identified phases of star formation. They are centrally condensed,  likely to be gravitationally bound and destined to ultimately form stars or clusters. They have power-law density profiles with flat inner regions, implying that no protostar has yet formed in their center  \cite[e.g.][]{War1999, Bac2000}. Their linewidths are thermalized, implying no significant turbulence support \cite[e.g.][]{Mye1983, Bar1998, Goo1998, Kir2007}.  They show no evidence for substantial 
amounts of rotation \citep{Goo1993,Bar1998,Cas2002}.  Intensity ratios of different transitions show that starless cores are almost isothermal, with $T\sim 10$ K \citep{Taf1998,Taf2002,Taf2004}, although there may be small variations of the order of a few K \cite[e.g.][]{Eva2001}.  Their evolution depends critically on the physical conditions of the star forming region in which they are embedded. In addition, their relatively quiescent dynamical state allows for an easier interpretation of observations because of the absence of the effects of thermal and dynamical feedback from a central protostar. For these reasons prestellar cores are ideal probes of the ultimate stellar origins: the initial conditions of star formation. 

Prestellar cores are observed to be magnetic \cite[see, e.g.][for a review]{Hei2005}. The magnetic field can provide support against the self-gravity of a molecular cloud or core, and thus affect its dynamical evolution. The amount of magnetic support is quantified by the mass to magnetic flux ratio $M/\Phi_B$ of the object under consideration. There is a critical value for the mass-to-flux ratio, 
\beq
\left(\frac{M}{\Phi_B}\right)_{\rm crit} = \left(\frac{1}{63G}\right)^{1/2}
\eeq
\citep{Mou1976} where $G$ is the gravitational constant. If the mass-to-flux ratio of an object exceeds this value (the object is magnetically supercritical), then the magnetic field is not strong enough to support the object against its own self-gravity, and the object contracts dynamically. The opposite is true for mass-to-flux ratios below the critical value (magnetically subcritical objects). Magnetic models of cloud fragmentation and core collapse predict that prestellar cores are magnetically {\em supercritical}, dynamically collapsing fragments, formed in magnetic parent clouds. These parent clouds which can be magnetically subcritical as a whole, and the supercritical fragments are formed through the process of ambipolar diffusion. The latter is the process of neutral particle diffusion through ions and magnetic field lines toward centers of gravity which increases the mass-to-flux ratio of the fragment \cite[e.g.][]{Fie1993}. 

Determining the amount of magnetic support in a molecular cloud as a whole and in individual molecular cloud cores is made complicated by at least the following three effects.

First of all,  Zeeman measurements only trace the component of the magnetic field along the line of sight, and for this reason geometrical considerations due to unknown cloud and core orientations limit the amount of information that can be obtained on an object-to-object basis \cite[e.g.][]{Shu1999,Tro2008,MT09}.

Second, in the case of Zeeman measurements, the magnetic field strength is convolved with the abundance profile of the tracer molecule, and, as a result, it is non-trivial to determine which parts of a core contribute most to a finite-beam Zeeman measurement of its magnetic field. Understanding the time evolution and spatial variation of the abundance of different species is  essential. Observations show that when such measurements are made 
using different molecular lines, they reveal different values of the magnetic field which cannot be reconciled. 
A characteristic example is CN and OH Zeeman observations of the same objects yielding different results \cite[e.g.][]{Fal2008}. Similarly, \citet{Cru2000} have argued, in the case of L1544, that OH data (and the associated Zeeman measurement of the magnetic field) do not sample the small, dense core observed in dust emission, so \citet{Cru2004} concluded that their Chandrasekhar-Fermi--measured plane-of-sky magnetic field may be discrepant from the Zeeman-measured  line-of-sight magnetic field of \citet{Cru2000}. This may be a result not only of geometrical projection effects, but also because different measurements correspond to different parts of the core. To verify the origin of such discrepancies, we need to understand how different molecules trace different parts of cores and thus different magnetic field values. This becomes especially important when attempting to study the variation of the magnetic field strength with density \cite[e.g.][]{Cru2010}.

Third, by the time the column density contrast between core and cloud becomes large enough for the core to be detected with high statistical significance and to be studied in detail, the core has already left the quasistatic contraction phase and has entered the phase of dynamical collapse, even in models which are originally heavily magnetically supported \citep{TM04}. At this stage, the mass-to-flux ratio is already larger than critical regardless of its value at larger scales \cite[e.g.][and references therein]{MTK06}. At even higher central densities, during the very advanced stages of collapse, the dynamical evolution of cores and the resulting density and velocity profiles have been shown to be quite insensitive to the initial value of the mass-to-flux ratio \citep{TM07b}.

In addition to the mass-to-flux ratio, the effect of the magnetic field on the dynamics of molecular clouds and molecular cloud cores also depends on the degree of ionization of the cloud or core. The ambipolar diffusion timescale is proportional to the degree of ionization, 
\begin{equation}
\tau_{\rm AD} \propto \frac{n_i}{n_{\rm H_2}}
\end{equation}
\citep{Cio1993}. 
This dependence
is physically straight-forward to understand: a higher degree of ionization implies a better coupling between magnetic field and matter, which results in an increased resistance of the ions as the neutrals drift past them and consequently an increased ambipolar diffusion timescale. 
The degree of ionization of a cloud or core is determined by the ionization rate (in dense cores the dominant ionization mechanism is cosmic ray ionization, and hence the relevant quantity is the cosmic ray ionization rate $\zeta$) and by the  relevant recombination reactions. Because of its feedback effect on the dynamics, and especially in the case where the chemistry is out of equilibrium, the degree of ionization is non-trivial to obtain; ideally, it has to be self-consistently calculated through a model following both dynamics and chemistry simultaneously.

For these reasons, the role of magnetic fields in the fragmentation of molecular clouds and the core formation and evolution process remains observationally uncertain and a hotly debated subject in the field.

In a companion paper \cite[][hereafter Paper I]{PaperI} we discussed our models of evolving prestellar cores which couple non-equlibrium chemistry with dynamics. Our extensive parameter study included magnetic and non-magnetic dynamical models with varying initial values of the mass-to-flux ratio or collapse retardation times respectively, varying C/O ratios, cosmic-ray ionization rates, and temperatures. In Paper I we focused on the evolution of the abundances of the most abundant and commonly observed molecules, and their dependence on the various model parameters. 

Here, we focus on the properties of the model cores that can help elucidate the role of magnetic fields in core formation and evolution: the degree of ionization, the most common ions and the mean ion molecular weight, abundance profiles of molecules used in Zeeman observations, and how the magnetic field that would be measured through Zeeman observations in our model cores depends on the actual strength of the magnetic field at the core center.

This paper is organized as follows. A brief review of the different models we consider is given in \S \ref{mod}. The results and their dependence on the various parameters we have studied are presented in \S \ref{res}. We summarize and discuss our conclusions in \S \ref{disc}.

\section{Models}\label{mod}

\begin{figure}
\plotone{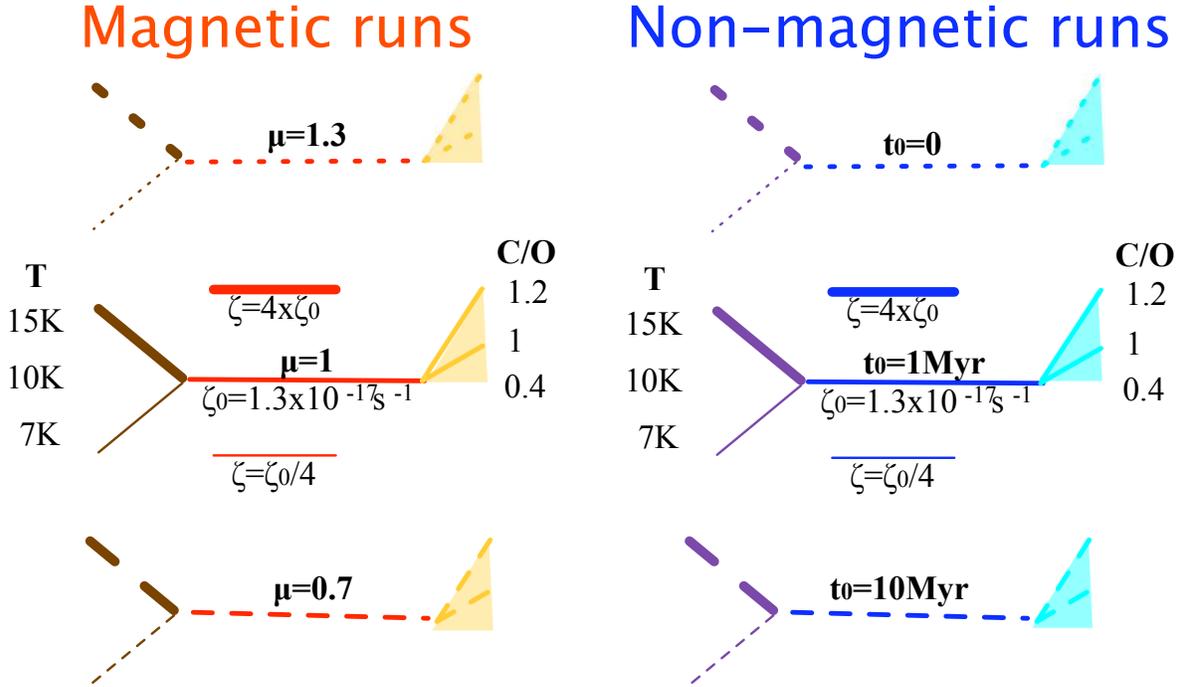}
\caption{\label{ModVis} 
Line types and colors used to denote each of the models studied in this work, unless otherwise noted. Solid normal-thickness red line: ``reference'' magnetic model; solid normal-thickness blue line: ``reference'' non-magnetic model. Dotted lines: ``fast'' models; dashed lines: ``slow'' models. Brown/purple lines: magnetic/nonmagnetic models with temperatures differing from the ``reference'' models. Orange/cyan shaded areas: variation in C/O ratio. Thin/thick solid red/blue lines: lower/higher cosmic-ray ionization rate magnetic/non-magnetic models (see text for details). }
\end{figure}

Since in this paper we are interested not only in the measurement of the magnetic fields but also in the evolution of the degree of ionization, we consider both magnetic and non-magnetic models. The details of the dynamical and chemical models of each class is discussed in detail in \S 2 and 3 of Paper I. Here, we very briefly review the parameters we have varied for each class of models presented and discuss in the following sections. These parameters include the temperature, the C/O ratio, the cosmic ray ionization rate, and a parameter 
controlling the time available for chemical evolution. The latter is the mass-to-flux ratio in the case of magnetic models, and the collapse delay time in the 
case of non-magnetic models (an initial time period during which chemistry evolves but the core does not 
evolve dynamically, representing an early stage of support due to turbulence which later decays).

Our ``reference'' magnetic model has a  mass-to-flux ratio equal to the critical value for collapse, a temperature of 10 K, a C/O ratio of 0.4, and a cosmic ray ionization rate of   $\zeta = 1.3 \times 10^{-17} {\rm \, s^{-1}}$.  
Our ``reference'' non-magnetic model has a collapse delay time (hereafter ``delay'') of 1Myr, and values for the C/O ratio, temperature, and $\zeta$ identical to those of the ``reference'' magnetic model. 
 
 For magnetic models
we examine two additional values of the initial mass to magnetic flux ratio: 1.3 times the critical value 
(a faster-evolving, magnetically supercritical model), and 0.7 of the critical value (a slower, 
magnetically subcritical model). For non-magnetic models we examine two additional values of delay: zero, and 10Myr. 

For each of these six dynamical models the carbon-to-oxygen ratio is varied from its ``reference'' value by keeping the abundance of C constant and changing that of O. The two other values of C/O ratio examined are 1 and 1.2. 
We have studied in this way a total of 18 different models (9 magnetic and 9 non-magnetic). 

In addition, to test the effect of the temperature, we have varied each of the six basic dynamical models by changing $T$ by a factor of  $\sim 1.5$ from its reference value of 10 K and examined models with $T = 7$ K and $T = 15$ K. We have thus studied 12 models (6 magnetic and 6 non-magnetic) with temperature varied from its reference value.

Finally, to test the effect of the cosmic ray ionization rate, we have studied four additional models (two magnetic and two non-magnetic), which have a ``reference'' value for the temperature, C/O ratio, and mass-to-flux ratio or delay (for magnetic and non-magnetic models respectively), but for which $\zeta$ is varied by a factor of four above ($\zeta = 5.2 \times 10^{-17}$ $s^{-1}$) and below ($\zeta = 3.3 \times 10^{-18}$ $s^{-1}$) its ``reference'' value (covering the range of observational 
estimates \cite[e.g.][]{McC2003,Hez2008}.

These additional  models bring the total of different models we have run and examined to 32. Figure \ref{ModVis} visually depicts these different models and the line type/color used to denote each one (unless explicitly noted otherwise).
\begin{figure*}
\plotone{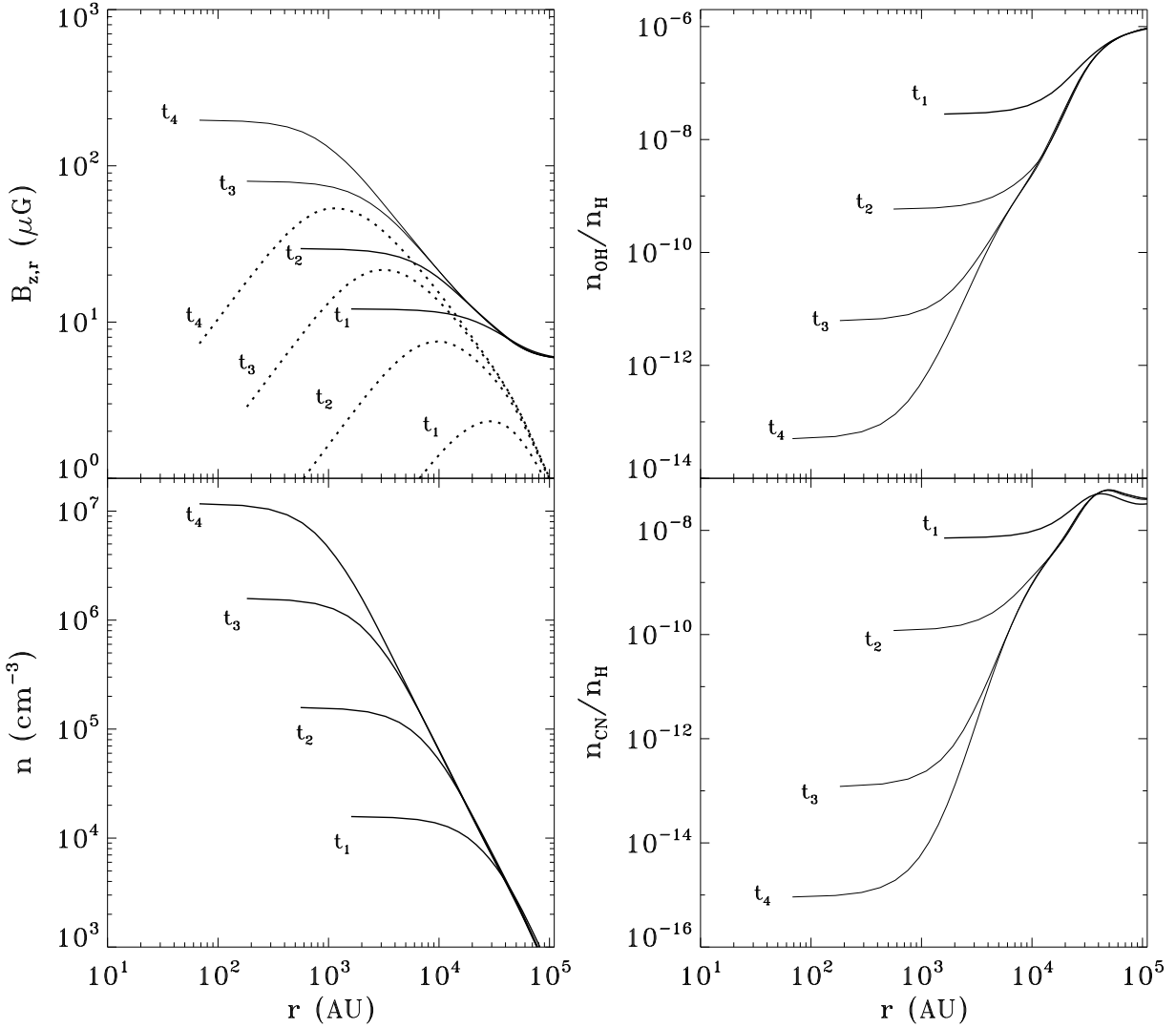}
\caption{\label{Bmulti} 
Left column: upper panel: radial profiles of the $z$-component (solid lines) and $r$-component (dotted lines) of the $B-field$ at four different times $t_1-t_4$ (corresponding to the same snapshots for which the average $B_z$ traced by OH and CN is shown in Fig.~\ref{meanBz}); lower panel: radial profiles of the number density at the same times $t_1-t_4$. 
Right column:  radial profiles of the OH (upper panel) and CN (lower panel) abundance taken at the same times as in the left column. 
}
\end{figure*}

\begin{figure*}
\epsscale{0.8}
\plotone{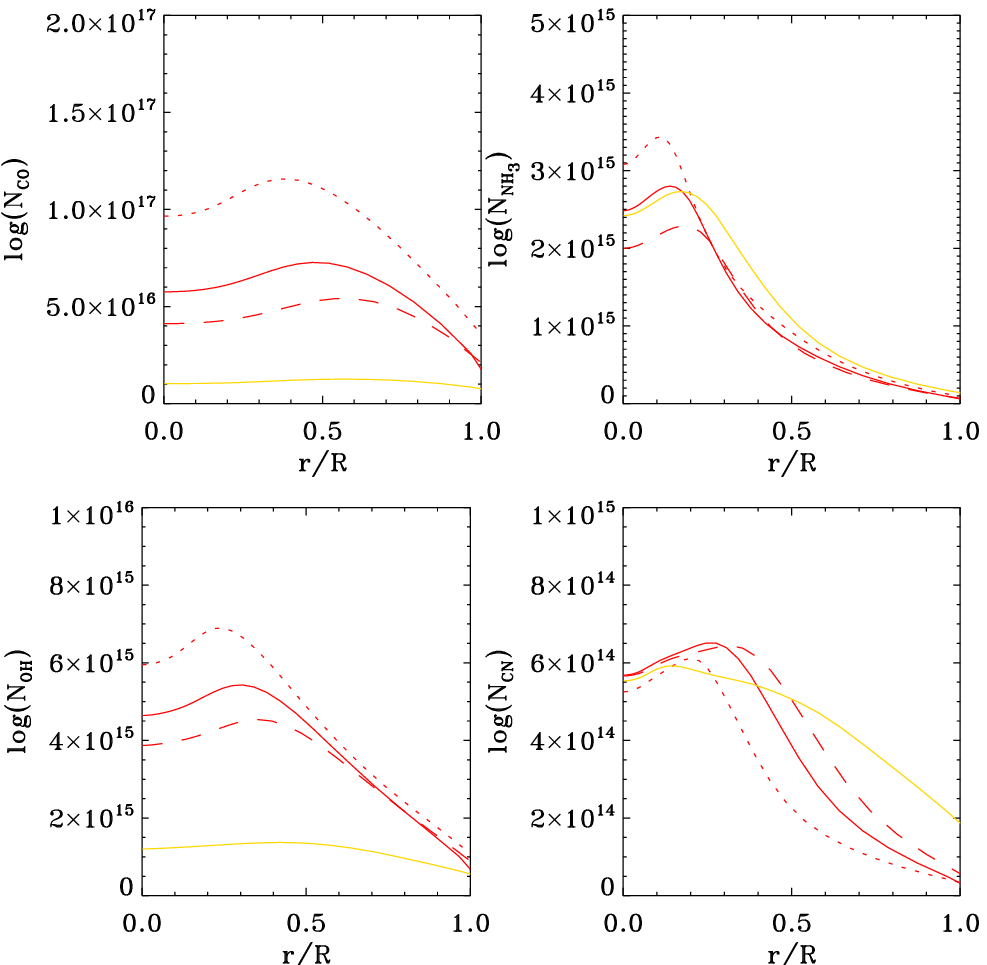}
\caption{\label{Colden} 
Column density of CO, NH$_3$, OH, and CN plotted against the 
fractional radius of the core, for a central density of $10^6 {\rm cm^{-3}}$  
(corresponding to snapshot $t_3$ in Fig.~\ref{Bmulti}). 
 The three red lines correspond to the different magnetic
dynamical models (magnetically critical, subcritical and supercritical
for the solid, dashed, and dotted lines respectively) 
for a reference value of the C/O ratio equal to 0.4, while the
yellow line corresponds to a magnetically critical model with C/O
ratio equal to 1 (see Fig.~\ref{ModVis}).}
\end{figure*}

\section{Results}\label{res}
\subsection{Zeeman-Traced Core Magnetic Field}

\begin{figure}
\plotone{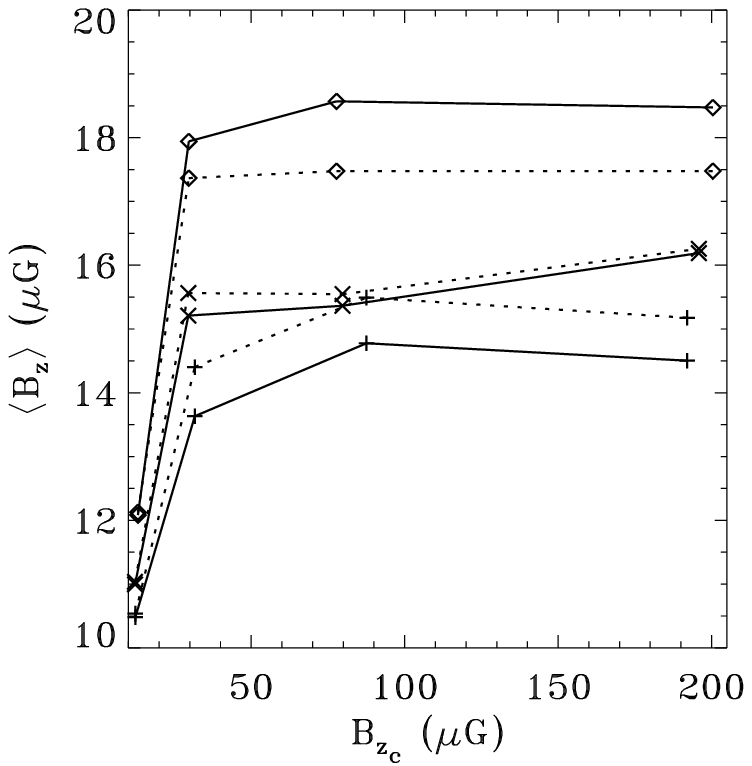}
\caption{\label{meanBz} 
Mean value of the z-component of the magnetic field of a core traced by
OH (solid lines) and CN (dotted lines) for the initially magnetically subcritical
(diamonds), critical ($\times$), and supercritical ($+$) cloud models.
For comparison, the initial magnetic field of the model cloud is 
$B_z=$4.2, $B_z=$5.6, and $B_z=$7.5 for the magnetically supercritical, 
critical and subcritical cloud models, respectively.
}
\end{figure}

\begin{figure*}
\plotone{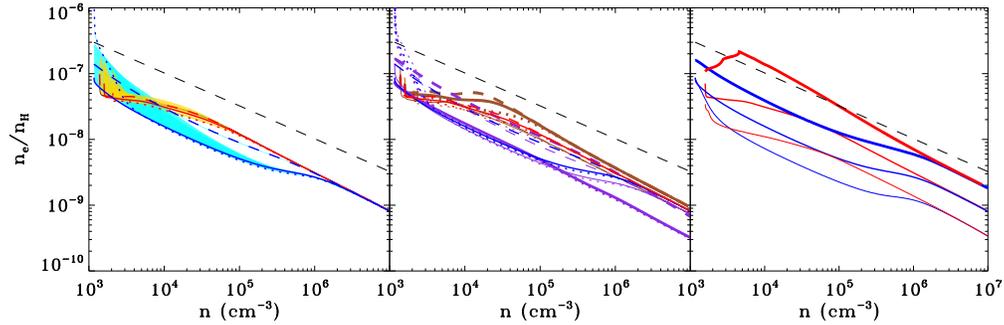}
\caption{\label{degion} 
Evolution of central electron abundance versus central number density
in magnetic and non-magnetic models. In all panels, red lines correspond to magnetic 
models with reference values for C/O  and temperature (dashed: magnetically subcritical; solid: magnetically critical; dotted: magnetically supercritical), 
and blue lines to non-magnetic models with reference values for C/O  and temperature  (dashed: 10Myr delay; solid: 1Myr delay; dotted: no delay). The black dashed line corresponds to Eq.\ (\ref{canonical}).
Left panel: effect of  a varying C/O ratio. The cyan and orange shaded areas (for non-magnetic and magnetic models respectively) correspond to a range of C/O values between $0.4$ (reference value) and $1.2$.
Middle panel: effect of a varying core temperature. The thin and thick brown (purple) solid lines correspond to temperature of 7 K and 15 K respectively for magnetic (non-magnetic) models; the ``reference'' value for the temperature is 10 K.  
Right panel: effect of a varying cosmic-ray ionization rate. The thick and thin red (blue) solid lines correspond to $\zeta$ a factor of four above and below the ``reference'' value for magnetic (non-magnetic) models (see Fig.~\ref{ModVis}).  }
\end{figure*}

In this section we examine how the magnetic field value that is measured through finite-beam Zeeman observations compares to the magnetic field at the center of a core. We consider two molecules commonly used for Zeeman observations, OH and CN.
Figure \ref{Bmulti} shows radial profiles of the magnetic field (left column, upper panel; $z-$ and $r-$components are shown with solid and dotted lines respectively), the OH abundance (right column, upper panel) and the CN abundance (right column, lower panel) at four time instances $t_1-t_4$, each of which represents an increase in the H$_2$ number density at the center of the core of an order of magnitude.  The central number density profiles at these times are shown in the lower panel of the left column. These radial profiles correspond to the magnetically critical (``reference'') dynamical model. Note that although CN and OH appear significantly depleted at the very central parts of the core, these results are consistent with observations by \citet{HBetal10}, who find that significant amounts of
CN remain at densities of $\sim 3\times 10^4 {\rm \, cm^{-3}}$, where CO has already
depleted. Comparing with the lower-left panel of Fig.~\ref{Bmulti}, we find that significant depletion of CN (abundance more than an order of magnitude smaller than in the outskirts of the core)  only sets in at densities higher than $\sim 10^5 {\rm \, cm^{-3}}$. To facilitate comparison with such observations, we show, in Fig.~\ref{Colden}, the column density of four molecules of interest (CO, NH$_3$, OH, and CN) as a function of (linearly plotted) fractional radius of
the core, for a central density of 10$^6$ cm$^{-3}$, corresponding to
snapshot $t_3$ in Fig.~\ref{Bmulti}, when the central abundance of CN for example
is more than five orders of magnitude lower than that in the outskirts
of the core. The three red lines correspond to the different magnetic
dynamical models (magnetically critical, subcritical and supercritical
for the solid, dashed, and dotted lines respectively) for C/O ratio equal to 0.4, while the
yellow line corresponds to a magnetically critical model with C/O
ratio equal to 1.  A peak at relatively large radius and then a plateau/mild
 decrease of the column density toward the center is a sign of
 depletion. In this context,  CO and NH$_3$ represent extreme cases of significant and little depletion respectively. 
OH and CN are intermediate cases, 
 both peaked at smaller radii than CO, in agreement with the findings
 of \citet{HBetal10}. If C/O is 1 rather than the
 fiducial 0.4 then CN exhibits a central decrease/plateau at much smaller
 radii, while CO and OH are significantly depleted.

In Fig.~\ref{meanBz} the mean magnetic field that would be seen through Zeeman measurements by a 0.1pc beam (extending out to a radius where the neutral number density falls to about $10^4 \, {\rm cm^{-3}}$) is plotted against the actual central value of the $z-$component of the magnetic field. The mock Zeeman-obtained ``observed'' value of the field is derived as follows. We calculate the average $z$-component magnetic field in the magnetic core under consideration, weighted by the number density of the corresponding molecule (OH or CN respectively) within a radial extent of 0.1 pc (the assumed size of the beam) at four different time instances.  The core is assumed to be viewed face-on (down the $z-axis$). In this way, geometrical effects do not enter our calculation of the mock Zeeman observation (the $z-$component of the magnetic field, which we examine here, is oriented exactly along the line of sight), and we can instead focus only on the effect of depletion. 

We then compare each of the averages derived in this way to the value of the z-component of the magnetic field at the center of the core at the same time.  In practice, this means that $\langle B_z \rangle$ is calculated through weighting by the OH or the CN number density in each radial annulus within the beam, respectively, assuming optically thin lines. Three different dynamical models are shown: magnetically subcritical (diamonds), critical ($\times$) and supercritical ($+$). 
Solid lines are used for the magnetic field traced by OH and dashed lines for the magnetic field traced by CN.

In all cases we can see that, although at low central densities the value of the Zeeman-traced magnetic field is close to the actual value of the central $B_z$, at higher densities the measured field increases only mildly and shows an overall increase by only a factor of two, whereas the actual central value of $B_z$ has increased by an order of magnitude. Most of the increase in the Zeeman-measured field strength takes place at low densities, while at higher densities the measured field tends to saturate. The reason for this behavior is seen in Fig.~\ref{Bmulti}:  the abundance of OH and CN is falling at the center of the core much more rapidly with time than the density increases. As a result, the Zeeman-traced magnetic field corresponds to the outer layers of the core ``onion skin'', where OH and CN have still relatively higher abundances  

\subsection{Degree of ionization}

\begin{figure*}
\plotone{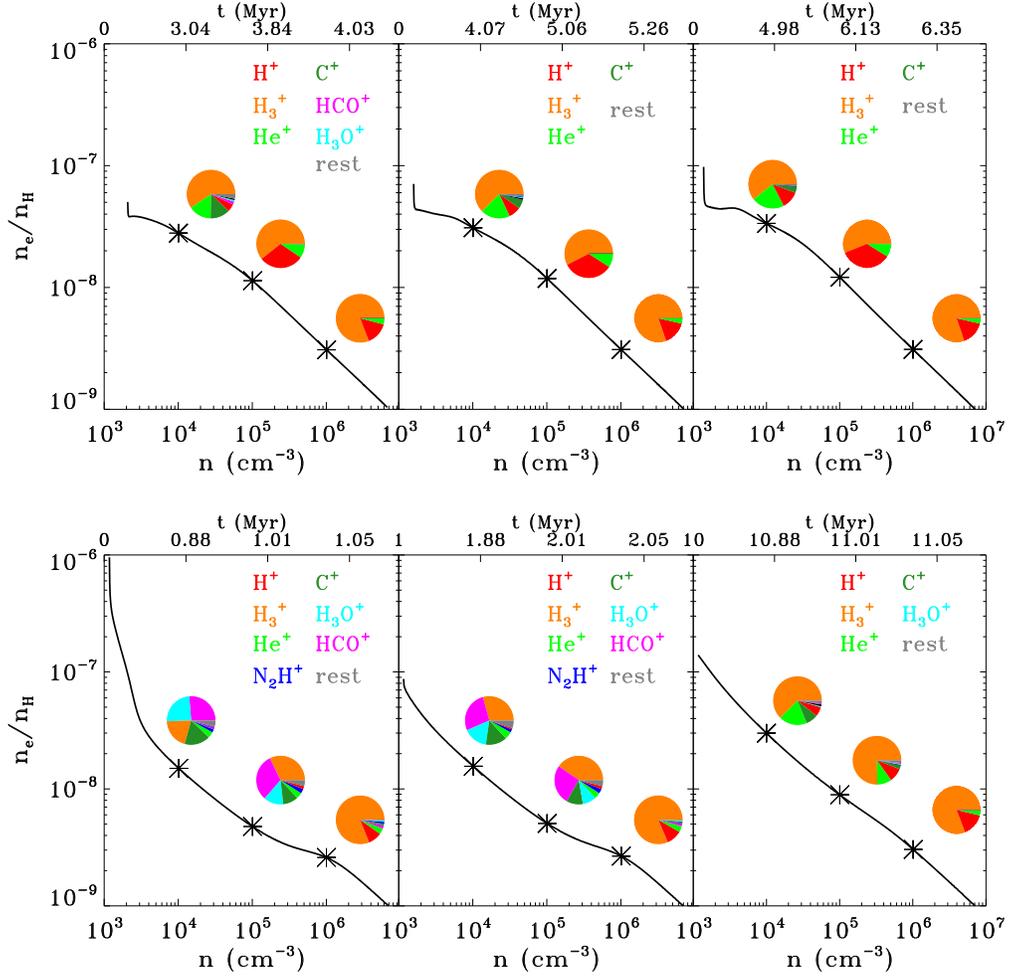}
\caption{\label{degion_pie} 
Degree of ionization as a function of central number density $n_c$ for the three magnetic (upper panels) and non-magnetic (lower panels) models, with ``reference'' values for the C/O ratio and the ionization rate $\zeta$.  Left column: ``fast'' models; middle column: ``reference'' models; right column: ``slow'' models.  Pie charts show the relative contribution of the dominant ions at the central densities denoted by stars. 
}
\end{figure*}

Figure \ref{degion} shows the evolution of electron abundance (degree of ionization) versus central number density in magnetic and non-magnetic models.  In all panels red lines correspond to magnetic models with reference values for C/O and temperature (dashed: magnetically subcritical; solid: magnetically critical; dotted: magnetically supercritical), 
and blue lines to non-magnetic models with reference values for C/O  and temperature (dashed: 10Myr delay; solid: 1Myr delay; dotted: no delay; see also Fig.~\ref{ModVis}). The left panel shows the effect of varying the C/O ratio, the middle panel the effect of varying the temperature, and the right panel the effect of varying the degree of ionization.  
For comparison we overplot with the black dashed line the usually adopted scaling [obtained through fits to earlier calculations, e.g., \citet{Elm1979,Nak1979}] between degree of ionization and number density, 
\begin{equation}\label{canonical}
\frac{n_i}{n} =  K_0\left(\frac{n}{10^5{\rm cm^{-3}}}\right)^{k-1}
\end{equation}
\citep{BM94}, with $k=0.5$ and $K_0 = 3\times 10^{-8}$.

It is obvious from Fig.\ \ref{degion} that at low central densities the out-of-equilibrium chemistry causes a variety of transitional effects, and the scaling of the degree of ionization with density cannot be expressed in a simple form such as Eq. (\ref{canonical}). For central densities higher than $\sim 10^5 {\rm \, cm ^{-3}}$ for magnetic models and $\sim 10^6 {\rm \, cm^{-3}}$ for non-magnetic models the scaling does asymptotically approach a simple power-law. The magnetic models approach power-law scaling at lower central densities because of the increased time available to the chemistry to overcome transitional effects before a specific density is reached.   The slope of this scaling (-0.6, corresponding to $k=0.4$) is steeper than the value usually adopted, but well within the uncertainties quoted in \citet{BM94}, who give a range for $k$ between $0.3$ and $0.5$. Our value for the slope is also consistent with the findings of \citet{Cas2002a} for the case of L1544. The normalization of our models is also consistently lower than Eq.\ (\ref{canonical}), but the discrepancy is not larger than the uncertainty on $K_0$ (an order of magnitude).

Except for the earliest evolutionary times that are dominated by transitional effects in chemistry, the slope of the scaling initially steepens (the value of $k$ in Eq.\ \ref{canonical} decreases) with increasing density, in agreement with \citet{Cio1994}. However, once $k$ reaches 0.4, the scaling of the degree of ionization with density becomes a power law, in contrast to \citet{Cio1994}, who found that $k$ continues to decrease and the scaling continues to steepen. This is likely a result of the (equilibrium) chemistry network adopted by \citet{Cio1994}, which was considerably more simplified than the (non-equilibrium) chemical model we use here, and which, as we will see below, resulted in a considerably different ion population, dominated by different species. 

The non-magnetic models show a qualitatively different behavior, with the scaling of the degree of ionization with density becoming more shallow before it later steepens. At sufficiently high central densities the non-magnetic models asymptotically approach the same power laws as the magnetic models with the same value of $\zeta$. The normalization of the scaling changes with $\zeta$ approximately as $\zeta^{1/2}$, consistent with the usually assumed scaling \cite[see, e.g., ][Eq.\ 24]{McK2007}.
The dependence at high densities on temperature is small (the asymptotic normalization being $\propto T^{1/2}$), however at low densities it can be considerable. Similarly, the dependence on the C/O ratio can be appreciable at low densities.

At sufficiently late evolutionary times and their associated high central densities magnetic and non-magnetic models with the same value of $\zeta$ and $T$ converge to the same scaling, regardless of delay time, C/O ratio, or mass-to-flux ratio. We note that the high-temperature non-magnetic model has not yet converged to its asymptotic form by the highest density displayed in Fig. \ref{degion}; for this reason, we have extended this run to even higher central densities, and we have confirmed that above $10^7 {\rm cm^{-3}}$ the degree of ionization does indeed approach the corresponding scaling of the magnetic model with the same values of $\zeta$ and $T$. 

\begin{figure*}
\plotone{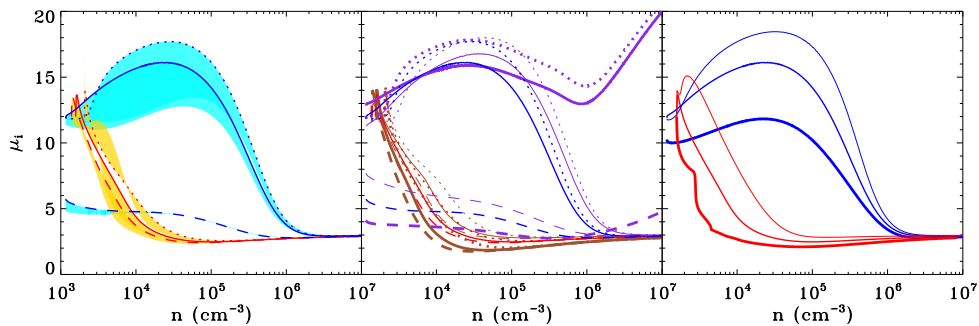}
\caption{\label{muion} 
Mean molecular weight $\mu_i$ at the center of the core, plotted as a function of central number density $n_c$. Panels, lines, and colors as in Fig.~\ref{degion} (see also Fig.~\ref{ModVis}).
}
\end{figure*}

\subsection{Dominant ions and mean molecular weight}

Figure \ref{degion_pie} shows the degree of ionization at the center of the core as a function of central number density $n$ for the three magnetic (upper panels) and non-magnetic (lower panels) models, with ``reference'' values for the C/O ratio, temperature, and cosmic-ray ionization rate $\zeta$.  ``Reference'' dynamical models are shown in the middle column, with ``fast'' and ``slow'' models in the left and right columns, respectively. The pie charts above the lines show the relative contribution of the dominant ions to the total ion population at the densities marked by stars. 

Due to our adopted initial condition, which assumes all C to be ionized, C$^+$ is initially the dominant ion. However, once the chemistry starts operating, this evolves to a different mixture.  Non-magnetic models have a more diverse mixture of ions at early times, with significant contributions from HCO$^+$ and H$_3$O$^+$.

At late evolutionary times H$_3^+$ is the dominant ion in all models, with  H$^+$ and He$^+$ being the most important secondary contributors (grain growth, not considered here, may alter the relative importance of  H$_3^+$ and H$^+$ in favor of H$^+$, see \citealp{Fl2005}).  This is an important difference from previous studies \cite[e.g.][]{Cio1993,Des2001,TM2007,McK2010}, which have assumed that the ion population at comparable densities is dominated by HCO$^+$ or by metallic ions such as Na$^+$, both of which have much higher molecular weight, implying that the ion fluid has much higher inertia at a given number density. Interestingly, the highest-density pie chart for all magnetic models and the ``slow'' non-magnetic model are very similar, so this ion distribution appears to be the one to which the ionization profile of the core settles at late enough times. 

For magnetic models the ion with the next highest contribution at low densities is  C$^+$. Non-magnetic models, on the other hand, feature  additional substantially contributing ions, such as C$_3$H$_3^+$, H$_3$O$^+$, HCO$^+$.


Figure \ref{muion} shows the evolution with central number density of the mean molecular weight of ions $\mu_i$. Panels, lines and colors correspond to various magnetic and non-magnetic models as described in Fig.~\ref{degion} (see  also Fig.~\ref{ModVis}). 

As expected from the results of Fig.\ \ref{degion_pie}, contrary to the usually adopted values of $\mu_i$ which are in the range of 20-30, the mean molecular weight of ions generally stays below 20. At high densities $\mu_i$ asymptotically approaches 3 for all cases except the high-temperature non-magnetic model, consistent with H$_3^+$ dominating the ions in these cases (see also \citealp{Cas2002a}).  The non-magnetic "slow" model appears significantly different (has a much lower mean molecular weight of ions at low densities) than all other models because before the gas starts to collapse, the chemistry has already had 10Myr to evolve, and the mixture of ions is already
close to the characteristic "late time" mix of the other models. 

The low values of $\mu_i$ (for most of the central number density range) in the magnetic models implies that the ambipolar diffusion timescale is {\em shorter} than the one computed using the usually adopted higher value for $\mu_i$, due to the smaller inertia of the ion fluid (which is attached to the magnetic field lines at these densities) and its associated decreased ability to provide resistance to the diffusion of the neutral fluid towards centers of gravity. 

The C/O ratio is shown to affect  the mean molecular weight appreciably. Its effect is stronger in the non-magnetic models, since oxygen and carbon-bearing molecules are more abundant in the gas phase in faster-evolving models such as the ``reference'' and ``fast'' nonmagnetic models. Indeed, the ``slow'' nonmagnetic model shows a much smaller sensitivity to the C/O ratio, comparable to that of the magnetic models. Similarly, the temperature has a significant effect on $\mu_i$, mainly through its effect on the chemistry of N-bearing molecules (see also discussion in Paper I). Specifically, we have verified that the increase in the mean molecular weight with temperature in the non-magnetic model (which persists even at high densities) is due to an increased abundance of N$_2$H$^+$. By contrast in the case of the magnetic high-temperature models,  H$^+$  is the dominant ion at intermediate central densities and $\mu_i$ decreases below three.

\section{Conclusions}\label{disc}

We have examined the effect that non-equilibrium chemistry in dynamical models of collapsing cores has on molecular abundances and thus on measurements of the magnetic field in these cores, on the degree of ionization and on the mean molecular weight of ions. We have considered both magnetic and non-magnetic models and models with different C/O ratios, cosmic-ray ionization rates, and core temperatures. 

We have found that molecules usually used in Zeeman observations of the line-of-sight magnetic field (OH and CN) have an abundance that decreases toward the center of the core much faster than the density increases (see Fig.\ \ref{Bmulti}). Thus, Zeeman measurements tend to sample the outer layers of the core and consistently underestimate the core magnetic field, especially for higher-density cores with higher magnetic field strengths. 

The degree of ionization was found to follow a complicated dependence on the number density in early evolutionary phases, when central densities are less than $10^5 \, {\rm cm^{-3}}$ for magnetic models and $10^6 \, {\rm cm ^{-3}}$ for non-magnetic models. At higher densities the scaling approaches a power-law with a slope of -0.6 (slightly steeper than the usually assumed value of -0.5) and with a normalization that scales with the product of $\zeta$ and $T$ to the 1/2 power; however, we note that for non-magnetic models with temperatures higher than 10 K, the dependency does not attain its asymptotic form until much higher central densities. 

The mean molecular weight of the ions was found to be systematically lower than the usually assumed value of $20-30$, and, at high densities, to asymptotically approach a value of 3 for all models, due to the asymptotic dominance of H$_3^+$. The only exceptions are the 15 K non-magnetic models (no delay and 1Myr delay), for which N$_2$H$^+$ becomes dominant at higher densities due to the sensitivity of nitrogen chemistry to temperature (see paper I, \S 5.3). At low densities the dominating ions and the associated value of $\mu_i$ evolve with density, and exhibit sensitivity to the value of the C/O ratio, $\zeta$, and the temperature. The considerably lower value of $\mu_i$ for the magnetic models compared to the usually assumed one implies that ambipolar diffusion operates faster. 

\acknowledgements{We thank Paul Goldsmith and the anonymous referee for insightful and constructive comments that improved this paper. This work was carried out at the 
Jet Propulsion Laboratory, California Institute of Technology, under a contract with the National Aeronautics 
and Space Administration.  \copyright 2012. All rights reserved.}


\begin{thebibliography}{}
\bibitem[Bacmann et al.(2000)]{Bac2000} Bacmann, A., Andr{\'e}, P., Puget, J.-L., Abergel, A., Bontemps, S., \& Ward-Thompson, D.\ 2000, \aap, 361, 555. 

\bibitem[Barranco \& Goodman(1998)]{Bar1998}
 Barranco, J.~A., \& Goodman, A.~A.\ 1998, \apj, 504, 207. 
Coherent Dense Cores. I. NH 3 Observations.

\bibitem[Basu \& Mouschovias(1994)]{BM94}
Basu, S. \& Mouschovias, T. Ch.\ 1994, ApJ, 432, 720

\bibitem[Caselli et al.(2002a)]{Cas2002} 
Caselli, P., Benson, P.~J., Myers, P.~C., \& Tafalla, M.\ 2002a, \apj, 572, 238.

\bibitem[Caselli et al.(2002b)]{Cas2002a} Caselli, P., Walmsley, 
C.~M., Zucconi, A., et al.\ 2002b, \apj, 565, 344 

\bibitem[Ciolek \& Mouschovias(1993)]{Cio1993} 
Ciolek, G.~E., \& Mouschovias, T.~Ch.\ 1993, \apj, 418, 774 

\bibitem[Ciolek \& Mouschovias(1994)]{Cio1994} 
Ciolek, G.~E., \& Mouschovias, T.~C.\ 1994, \apj, 425, 142 

\bibitem[Crutcher \& Troland(2000)]{Cru2000} 
Crutcher, R.~M., \& Troland, T.~H.\ 2000, \apjl, 537, L139 

\bibitem[Crutcher et al.(2004)]{Cru2004} 
Crutcher, R.~M., Nutter, D.~J., Ward-Thompson, D., \& Kirk, J.~M.\ 2004, \apj, 600, 279 

\bibitem[Crutcher et al.(2010)]{Cru2010}
 Crutcher, R.~M., Wandelt, B., Heiles, C., Falgarone, E., \& Troland, T.~H.\ 2010, \apj, 725, 466 

\bibitem[Desch \& Mouschovias(2001)]{Des2001}
Desch, S.~J., \& Mouschovias, T.~Ch.\ 2001, \apj, 550, 314 

\bibitem[Elmegreen(1979)]{Elm1979} 
Elmegreen, B.~G.\ 1979, 
\apj, 232, 729 

\bibitem[Evans et al.(2001)]{Eva2001} 
Evans, N.~J., II, Rawlings, J.~M.~C., Shirley, Y.~L., \& Mundy, L.~G.\ 2001, \apj, 557, 193. 

\bibitem[Falgarone et al.(2008)]{Fal2008} 
Falgarone, E., Troland, T.~H., Crutcher, R.~M., \& Paubert, G.\ 2008, \aap, 487, 247. CN Zeeman measurements in star formation regions.

\bibitem[Fiedler \& Mouschovias(1993)]{Fie1993} 
Fiedler, R.~A., \& Mouschovias, T.~C.\ 1993, \apj, 415, 680.

\bibitem[Flower et al.(2005)]{Fl2005} 
Flower, D.~R., Pineau Des For{\^e}ts, G., \& Walmsley, C.~M.\ 2005, \aap, 436, 933 

\bibitem[Goodman et al.(1993)]{Goo1993} 
Goodman, A.~A., Benson, 
P.~J., Fuller, G.~A., \& Myers, P.~C.\ 1993, \apj, 406, 528.

\bibitem[Goodman et al.(1998)]{Goo1998} 
Goodman, A.~A., 
Barranco, J.~A., Wilner, D.~J., \& Heyer, M.~H.\ 1998, \apj, 504, 223.

\bibitem[Heiles \& Crutcher(2005)]{Hei2005} 
Heiles, C., \& Crutcher, R.\ 2005, Cosmic Magnetic Fields, 664, 137 

\bibitem[Hezareh et al.(2008)]{Hez2008}
 Hezareh, T., Houde, M., 
McCoey, C., Vastel, C., \& Peng, R.\ 2008, \apj, 684, 1221 

\bibitem[Hily-Blant et al.(2010)]{HBetal10} 
Hily-Blant, P., Walmsley, M., Pineau Des For{\^e}ts, G., \& Flower, D.\ 2010, \aap, 513, A41 

\bibitem[Kirk et al.(2007)]{Kir2007} 
Kirk, H., Johnstone, D., 
\& Tafalla, M.\ 2007, \apj, 668, 1042.

\bibitem[McCall et al.(2003)]{McC2003} 
McCall, B.~J., Huneycutt, A.~J., Saykally, R.~J., et al.\ 2003, \nat, 422, 500 

\bibitem[McKee \& Ostriker(2007)]{McK2007} McKee, C.~F., \& Ostriker, E.~C.\ 2007, \araa, 45, 565. 

\bibitem[McKee et al.(2010)]{McK2010} 
McKee, C.~F., Li, P.~S., \& Klein, R.~I.\ 2010, \apj, 720, 1612 

\bibitem[Mouschovias \& Spitzer(1976)]{Mou1976} 
Mouschovias, T.~C., \& Spitzer, L., Jr.\ 1976, \apj, 210, 326 

\bibitem[Mouschovias et al.(2006)]{MTK06} Mouschovias, T.~C., 
Tassis, K., \& Kunz, M.~W.\ 2006, \apj, 646, 1043 

\bibitem[Mouschovias \& Tassis(2009)]{MT09}
 Mouschovias, T.~C., \& Tassis, K.\ 2009, \mnras, 400, L15 

\bibitem[Myers(1983)]{Mye1983} 
Myers, P.~C.\ 1983, \apj, 270, 105. 

\bibitem[Nakano(1979)]{Nak1979} 
Nakano, T.\ 1979, \pasj, 31, 
697 

\bibitem[Shu et al.(1999)]{Shu1999} Shu, F.~H., Allen, A., 
Shang, H., Ostriker, E.~C., 
\& Li, Z.-Y.\ 1999, NATO ASIC Proc.~540: The Origin of Stars and Planetary Systems, 193 


\bibitem[Tafalla et al.(1998)]{Taf1998} 
Tafalla, M., Mardones, D., Myers, P.~C., Caselli, P., Bachiller, R., 
\& Benson, P.~J.\ 1998, \apj, 504, 900. 

\bibitem[Tafalla et al.(2002)]{Taf2002} 
Tafalla, M., Myers, 
P.~C., Caselli, P., Walmsley, C.~M., \& Comito, C.\ 2002, \apj, 569, 815.

\bibitem[Tafalla et al.(2004)]{Taf2004} 
Tafalla, M., Myers, P.~C., Caselli, P., \& Walmsley, C.~M.\ 2004, \aap, 416, 191. 

\bibitem[Tassis \& Mouschovias(2004)]{TM04} 
Tassis, K., \& Mouschovias, T.~C.\ 2004, \apj, 616, 283 

\bibitem[Tassis \& Mouschovias(2007)]{TM2007} 
Tassis, K., \& Mouschovias, T.~C.\ 2007a, \apj, 660, 388. 

\bibitem[Tassis \& Mouschovias(2007b)]{TM07b} 
Tassis, K., \& Mouschovias, T.~C.\ 2007b, \apj, 660, 402.

\bibitem[Tassis et al.(2011)]{PaperI} 
Tassis, K., Willacy, K., Yorke, H.W., \& Turner, N. 2011, {\it submitted to ApJ} (Paper I).

\bibitem[Troland \& Crutcher(2008)]{Tro2008} 
Troland, T.~H., \& Crutcher, R.~M.\ 2008, \apj, 680, 457 

\bibitem[van Leer (1979)]{vL79}
van Leer, B.\ 1979, J. Comput. Phys., 32, 101

\bibitem[Ward-Thompson et al.(1999)]{War1999} Ward-Thompson, 
D., Motte, F., \& Andre, P.\ 1999, \mnras, 305, 143. 


\end{thebibliography}
\end{document}